\documentclass[aps,prb, twocolumn, showpacs, reprint, floatfix]{revtex4-1}

\usepackage{graphicx}
\usepackage{color}
\usepackage{amsmath}
\usepackage{amssymb}
\usepackage{amsthm}
\usepackage{stmaryrd}
\usepackage{soul}
\usepackage{ulem}
\newcommand{\stkout}[1]{\ifmmode\text{\sout{\ensuremath{#1}}}\else\sout{#1}\fi}

\newcommand{\mathsym}[1]{{}}

\begin{document}

\title{Spin-resolved orbital magnetization in Rashba two-dimensional electron gas}
\author{A. Dyrda\l$^{1}$, V. K.~Dugaev$^{2}$, and  J.~Barna\'s$^{1,3}$ }
\address{
$^1$Faculty of Physics, Adam Mickiewicz University,
ul. Umultowska 85, 61-614 Pozna\'n, Poland \\
$^2$Department of Physics and Medical Engineering, Rzesz\'ow University of Technology,
al. Powsta\'nc\'ow Warszawy 6, 35-959 Rzesz\'ow, Poland \\
$^3$Institute of Molecular Physics, Polish Academy of Sciences,
ul. M. Smoluchowskiego 17, 60-179 Pozna\'n, Poland }

\date{\today }

\begin{abstract}
We calculate orbital spin-dependent magnetization in a two-dimensional electron gas with spin-orbit interaction of Rashba type.
Such an orbital magnetization is admitted by the time-reversal symmetry of the system, and gives rise to spin currents when the system is not in thermal equilibrium. The theoretical approach is based on the linear response theory and the Matsubara Green's function formalism. To account for the spin-resolved  orbital magnetization a spin-dependent  vector potential has been introduced. The  spin currents which appear in thermal nonequilibrium due to the spin-resolved orbital magnetization play an important role in the spin Nernst effect, and have to be included in order to  correctly describe the low-temperature spin Nernst conductivity.
\end{abstract}

\maketitle

\section{Introduction}
Heat currents coupled to electric and spin currents can be effectively used in novel spintronics devices to control
not only charge and entropy/energy transport but also to control spin transport. The latter is the main goal of spin caloritronics -- a new branch of spin electronics. Indeed, there is currently huge interest, both experimental and theoretical, in thermal generation of spin currents which in turn can be used to control magnetic state of a system. One should mention here the  Seebeck and spin Seebeck effects, the Nernst and spin Nernst effects, and also others.

Theoretical description of the phenomena that occur as a system's response to a temperature gradient is generally more complex than description of a system subject to an external electric field.  To describe heat/energy transport in the  framework of Green function formalism and Kubo formula, an auxiliary vector potential has been introduced\cite{Ma,Dyrdal2013,Tatara2015_1,Tatara2015_2} instead of the Luttinger 'gravitational' potential introduced earlier~\cite{Luttinger}. Such a  vector potential may be considered as an analogue of the vector potential in the theory of electromagnetism.

It is well known in the relevant literature, that to determine the system's response
(namely, the transverse electric current)
to a temperature gradient, the orbital magnetization should be taken into account in order to get results that obey the fundamental  thermodynamics laws. In other words, electric current due to a nonzero orbital magnetization ensures physical behavior of the off-diagonal kinetic coefficients in the zero-temperature limit, such as off-diagonal electrical conductivity due to thermal bias. This problem was studied first by Obraztsov~\cite{obraztsov}, who  introduced the magnetization currents to the problem of the off-diagonal thermal transport in order to satisfy the Onsager relations of the kinetic coefficients. This problem has been then raised in many papers, e.g. in the context of quantum Hall effect~\cite{Streda_1977,Grivin,Streda_1985,QinNiuShi} or Nernst effect in fluctuating superconductors~\cite{Ruzin,Niu_2006} and graphene-like materials~\cite{Varlamov2014,Varlamov2015}.

The orbital magnetization appears as a consequence of the orbital motion of electrons when the time reversal symmetry in a system is broken~\cite{hirst1997,Niu2005,resta2005,resta2006}. This happens in the case of ferro- and ferrimagnets, or in nonmagnetic materials in an external magnetic field. In the presence of spin-orbit interaction, electron motion in a system is even more complex. Such an interaction can appear as an effective momentum-dependent magnetic field, and may lead to such phenomena like spin Hall and spin Nernst effects that require neither magnetic materials nor external magnetic fields.

An important question which arises in the context of spin-orbit interaction concerns behavior of the off-diagonal spin-kinetic coefficients in systems with time-reversal symmetry. An important example of such systems is the two-dimensional electron gas with Rashba spin-orbit interaction, that appears at the interface of semiconductor heterostructures. Thermal properties of such  systems have been studied recently in a couple of publications~\cite{Ma,Borge,Tolle,Gorini,Dyrdal2016}. In our recent paper we have shown that to describe properly the low-temperature behavior of the spin Nernst effect in a two-dimensional electron gas in frame of the linear response theory, one needs to introduce orbital effects as well -- even though  the system is symmetrical with respect to the time-reversal.
The usual orbital magnetization is then suppressed due to the time-reversal symmetry. Therefore, we have introduced the {\textit{spin-resolved orbital magnetization}} and have shown that it contributes to the spin current in thermal nonequilibrium, and therefore also to the spin Nernst conductivity~\cite{Dyrdal2016}.

In this paper we present detailed calculation of the spin-resolved orbital magnetization. In Sec.~2 we describe the model and also present symmetry arguments for the spin-resolved orbital magnetization. In Sec.~3 we introduce spin vector potential and calculate the relevant Green function, which is used in Sec.~4 to calculate the corresponding spin-dependent orbital magnetization. Results and discussion are presented in Secs.~5 and 6, respectively.

\section{Theoretical model}

Hamiltonian of a two-dimensional electron gas with Rashba spin-orbit interaction can be written  as
\begin{equation}
\label{1}
H_{R} = \frac{\hbar^{2} k^{2}}{2 m} \sigma_{0} + \alpha (k_{y}\, \sigma_{x} - k_{x}\, \sigma_{y}),
\end{equation}
where  $\sigma_{n}$ (for  $n = x, y, z$) are the Pauli matrices and $\sigma_{0}$ is the unit matrix. All these matrices operate in the spin space. The parameter $\alpha$ describes strength of the Rashba interaction, while $k_x$ and $k_y$ are the in-plane wavevector components. Eigenvalues of the above Hamiltonian have the form,  $E_{{\bf k}\pm} = \varepsilon_{k} \pm \alpha k$, with $\varepsilon_{k} = \hbar^{2} k^{2}/2 m$ and $k^{2} = k_{x}^{2} + k_{y}^{2}$.

The retarded Green's function corresponding to the Hamiltonian (1) can be written in the following form:
\begin{equation}
\label{9}
G_{\mathbf{k}}^{R}(\varepsilon) = G_{\mathbf{k} 0}^{R}(\varepsilon)\, \sigma_{0} + G_{\mathbf{k} x}^{R}(\varepsilon)\, \sigma_{x} + G_{\mathbf{k} y}^{R}(\varepsilon)\, \sigma_{y},
\end{equation}
where
\begin{subequations}
\begin{align}
\label{9a}
G_{ \mathbf{k} 0}^{R}(\varepsilon) = \frac{1}{2} [G_{ \mathbf{k} +}^{R}(\varepsilon) + G_{ \mathbf{k} -}^{R}(\varepsilon)],\\
\label{9b}
G_{ \mathbf{k} x}^{R}(\varepsilon) = \frac{1}{2} \sin(\phi) [G_{ \mathbf{k} +}^{R}(\varepsilon) - G_{ \mathbf{k} -}^{R}(\varepsilon)],\\
\label{9c}
G_{ \mathbf{k} y}^{R}(\varepsilon) = - \frac{1}{2} \cos(\phi) [G_{ \mathbf{k} +}^{R}(\varepsilon) - G_{ \mathbf{k} -}^{R}(\varepsilon)],
\end{align}
\end{subequations}
with $\phi$ denoting the angle between the wavevector $ \mathbf{k}$ and the axis $x$, and $G_{ \mathbf{k} \pm}^{R}(\varepsilon)$ defined as
\begin{eqnarray}
\label{10}
G_{ \mathbf{k} \pm}^{R}(\varepsilon) = \frac{1}{\varepsilon + \mu - E_{{\bf k}\pm} + i \Gamma}.
\end{eqnarray}
Here, $\Gamma$ is the imaginary part of the self energy, which is related to the appropriate relaxation time $\tau$, $\Gamma =\hbar/2\tau$.
The advanced Green's function can be written in a similar form with $\Gamma \to -\Gamma $.

The spin-orbit Rashba interaction is a consequence of a structural inversion asymmetry of the confinement potential in a quantum well.
This means that whenever $E_{\mathbf{k},\pm } = E_{-\mathbf{k},\pm }$ due to time-inversion symmetry,
$E_{\mathbf{k},+} \neq E_{\mathbf{k},-}$ due to the non-invariance with respect to spatial inversion.
The spin-orbit interaction can be then considered as a momentum-dependent magnetic field acting on the electron spin. However, this internal magnetic field does not break the  time-reversal symmetry, as also follows from the form of Hamiltonian (1), which is symmetrical with respect to time reversal. As a consequence of this symmetry, the orbital magnetization in the system is suppressed. However, time-reversal symmetry of the system under consideration allows for the spin-dependent orbital magnetization (or equivalently for spin dependent magnetic field $\mathbf{B}_{s}$), which has opposite orientation for spin-up and spin-down electrons.  The total orbital magnetization is then equal to zero, ${\bf M}={\bf M}_\uparrow +{\bf M}_\downarrow=0$, as ${\bf M}_\uparrow $ and ${\bf M}_\downarrow $ are oriented in the opposite directions, but the spin-resolved orbital magnetization defined as
${\bf M}_{orb}^{s}={\bf M}_\uparrow -{\bf M}_\downarrow$ is then nonzero, ${\bf M}_{orb}^{s}\ne 0$.

To calculate the spin-resolved orbital magnetization we introduce a spin vector potential,
${\bf A}_s({\bf r})=\sigma _z{\bf A}({\bf r})$, into the Hamiltonian (1) by the substitution
$- i \hbar \nabla \sigma_{0} \to -i \hbar \nabla\sigma_{0} -e\, {\bf A}_s$. This spin vector potential is related to the spin-dependent magnetic field $\mathbf{B}_{s} = \sigma_{z} \mathbf{B}$ according to the formula,  ${\bf B}_{s}={\rm rot}\, {\bf A}_{s}$. Thus, the effective spin-dependent magnetic field affects the orbital motion of spin-up and spin-down electrons in different ways. In the case considered here,  this spin magnetic field is oriented along the  $z$ axis (normal to the system's plane): $\mathbf{B} = (0, 0, B)$. The resulting Hamiltonian reads:
\begin{eqnarray}
\label{H_A}
H_{\mathbf{A}} = \frac{\hbar^{2}}{2m}
\Big( {\mathrm{\mathbf{k}}}\sigma_{0} - \frac{e}{\hbar} \mathbf{A}_{s}\Big) ^{2} + \alpha (k_{y} \sigma_{x} - k_{x} \sigma_{y}) \nonumber\\
- \alpha \frac{e}{\hbar}\, (A_{s y}\otimes \sigma_{x} - A_{s x} \otimes \sigma_{y}).
\end{eqnarray}

The main objective of the following sections is to calculate the total energy of the system in the presence of a nonzero $B$, and then to calculate the spin-resolved orbital magnetization as a derivative of this energy with respect to $B$, taken at
 $B\to 0$.

\section{Green's function}

The Green  function describing two-dimensional electron gas with Rashba interaction in the spin-dependent magnetic field $B_{s}$, see Eq.~(5),  satisfies the following equation written in the coordinate space:
\begin{widetext}
\begin{eqnarray}
\label{GF_eq1}
\int d^{2}{\mathbf{r}}' \left\{ \varepsilon + \frac{\hbar^{2} }{2m}
\left[ \nabla_{x}^{2} +\nabla_{y}^{2} - \frac{2ie}{\hbar} (A_{x} \nabla_{x} + A_{y} \nabla_{y})\right]
+ i\alpha  \Big[ \sigma_{x} \nabla_{y} - \sigma_{y} \nabla_{x}\Big]\right\}
\delta(\mathbf{r} - \mathbf{r}') \, \mathcal{G}(\varepsilon, \mathbf{r}', \mathbf{r}'')
 = \delta(\mathbf{r} - \mathbf{r}''), \hspace{0.3cm}
\end{eqnarray}
\end{widetext}
where we neglect the diamagnetic term proportional to $\textbf{A}_{s}^{2}$ and a contribution originating from the third term in Hamiltonian ({\ref{H_A}}), which gives a small correction since the Rashba interaction is assumed to be small. Note, that for notation brevity we write in this section $\varepsilon \equiv \varepsilon + \mu + i \delta \, \mathrm{sign}(\varepsilon)$ for the zero-temperature casual Green's function and $\varepsilon \equiv i \varepsilon_{n}$ for the Matsubara-Green's function.

Similarly as in the case of a constant magnetic field, we make use of the fact that the Green's function may be expressed as the product of the translationally and rotationally invariant core Green's function ${\mathcal{G}}_{0}(\varepsilon, {\mathbf{r}} - {\mathbf{r'}})$ and an exponential factor~\cite{Varlamov2014,Varlamov2015,ChenLee2011,Khodas},
\begin{equation}
\mathcal{G}(\varepsilon, \mathbf{r}, \mathbf{r}')
= {\mathcal{G}}_{0}(\varepsilon, {\mathbf{r}} - {\mathbf{r'}})\, {\rm{e}}^{i \mathcal{A}_{\mathbf{r} \mathbf{r'}} \sigma_{z}}\, ,
\end{equation}
where $\mathcal{A}_{\mathbf{r} \mathbf{r'}} \equiv \frac{e}{\hbar} \int_{\mathbf{r}}^{\mathbf{r}'} \mathbf{A}({\mathbf{R}})\cdot d{\mathbf{R}}$ is the Schwinger or Peierls phase factor. The integral  of gauge vector potential in this phase factor is along a straight line from $\mathbf{r}$ to $\mathbf{r}'$. Consequently, Eq.~(\ref{GF_eq1}) can be rewritten in the form
\begin{widetext}
\begin{eqnarray}
\label{GF_eq2}
\int d^{2}{\mathbf{r}}' \left\{ \varepsilon + \frac{\hbar^{2} }{2m}
\left[ \nabla_{x}^{2} + \nabla_{y}^{2} \right] + i\alpha  \left[ \sigma_{x}\nabla_{y} - \sigma_{y} \nabla_{x}\right]\right\}
{\rm e}^{i \sigma_{z} \mathcal{A}_{\mathbf{r} \mathbf{r}'}}
\delta(\mathbf{r} - \mathbf{r}')\, \mathcal{G}_{0}(\varepsilon, \mathbf{r}', \mathbf{r}''){\rm e}^{i \sigma_{z} \mathcal{A}_{\mathbf{r'} \mathbf{r}''}} = \delta(\mathbf{r} - \mathbf{r}'') {\rm e}^{i \sigma_{z} \mathcal{A}_{\mathbf{r} \mathbf{r}''}}.
\hspace{0.2cm}
\end{eqnarray}
\end{widetext}

We look for the Green's function ${\mathcal{G}}_{0}$ in the following form:
\begin{equation}
\label{gf}
\mathcal{G}_{0}(\varepsilon, \mathbf{r}'- \mathbf{r}'') = \sum_{i}
\mathcal{G}_{0 i}(\varepsilon, \mathbf{r}'- \mathbf{r}'') \,
\sigma_{i}
\end{equation}
where $i = \{0, x, y, z\}$. Thus, the equation (\ref{GF_eq2}) leads to a set of four equations for the four components of the Green function, $\mathcal{G}_{0 i}(\varepsilon, \mathbf{r} - \mathbf{r}')$.
Upon performing the Fourier transformation with respect to the space variables, this set of equations can be written in the following form (for details of calculations see Appendix A):
\begin{eqnarray}
\label{soeq}
\Lambda_{\mathbf{k}}(\varepsilon)  {\small{ \left(
          \begin{array}{c}
            \mathcal{G}_{\mathbf{k} 0}(\varepsilon) \\
             \mathcal{G}_{\mathbf{k} x}(\varepsilon)  \\
             \mathcal{G}_{\mathbf{k} y}(\varepsilon)  \\
             \mathcal{G}_{\mathbf{k} z}(\varepsilon)  \\
          \end{array}
        \right) = \left(
          \begin{array}{c}
            a_{\mathbf{k} 0}(\varepsilon) \\
             a_{\mathbf{k} x}(\varepsilon)  \\
             a_{\mathbf{k} y}(\varepsilon)  \\
             a_{\mathbf{k} z}(\varepsilon)  \\
          \end{array}
        \right)}},
\end{eqnarray}
where $\mathcal{G}_{\mathbf{k} i}(\varepsilon)$ (for $i = 0,x,y,z$) is the Fourier transform of
$\mathcal{G}_{0 i}(\varepsilon, \mathbf{r} - \mathbf{r}')$,
the matrix  $\hat{\Lambda}_{\mathbf{k}}(\varepsilon)$  is defined as
\begin{eqnarray}
\hat{\Lambda}_{\mathbf{k}}(\varepsilon) = {\small{\left(
		\begin{array}{cccc}
		[g_{\mathbf{k} 0}(\varepsilon)]^{-1} & - \alpha  k_{y} & \alpha k_{x} & 0 \\
		- \alpha k_{y} & [g_{\mathbf{k} 0}(\varepsilon)]^{-1} & 0 & i \alpha k_{x} \\
		\alpha k_{x} & 0 & [g_{\mathbf{k} 0}(\varepsilon)]^{-1} & i \alpha k_{y} \\
		0 & - i \alpha k_{x} & - i \alpha k_{y} & [g_{\mathbf{k} 0}(\varepsilon)]^{-1} \\
		\end{array}
		\right) }} ,\hskip0.7cm
\end{eqnarray}
with $[g_{\mathbf{k} 0}(\varepsilon)]^{-1} = \varepsilon - \varepsilon_{k}$, and the functions $a_{\mathbf{k} i}(\varepsilon)$ on the right hand side of Eq.(10) have the following form:
\begin{eqnarray}
a_{\mathbf{k} 0}(\varepsilon) = 1, \hspace{6.8cm}\\
a_{\mathbf{k} x}(\varepsilon) = - \alpha \frac{e}{2 \hbar} B  \partial_{k_{y}} G_{\mathbf{k} 0}(\varepsilon),\hspace{4.2cm}\\
a_{\mathbf{k} y}(\varepsilon) = \alpha \frac{e}{2 \hbar} B  \partial_{k_{x}} G_{\mathbf{k} 0}(\varepsilon),\hspace{4.5cm}\\
a_{\mathbf{k} z}(\varepsilon) = i \frac{e}{2 \hbar} B \left[  \partial_{k_{x}}[g_{\mathbf{k} 0}(\varepsilon)]^{-1} \partial_{k_{y}} G_{\mathbf{k} 0}(\varepsilon)\right. \hspace{2.3cm}\nonumber\\  -\left. \left(
\partial_{k_{y}} [g_{\mathbf{k} 0}(\varepsilon)]^{-1}\right) \left( \partial_{k_{x}} G_{\mathbf{k} 0}(\varepsilon)\right) \right].\hspace{0.9cm}
\end{eqnarray}
Thus, the core Green's function (\ref{gf}) in the momentum space takes the form
\begin{equation}
\label{gf_k}
\mathcal{G}_{\mathbf{k}}(\varepsilon) = \mathcal{G}_{\mathbf{k} 0}(\varepsilon) \sigma_{0} + \mathcal{G}_{\mathbf{k} x}(\varepsilon) \sigma_{x} + \mathcal{G}_{\mathbf{k} y}(\varepsilon) \sigma_{y}+ \mathcal{G}_{\mathbf{k} z}(\varepsilon) \sigma_{z} \, ,
\end{equation}
where $\mathcal{G}_{\mathbf{k} \alpha}(\varepsilon)$ are solutions of Eq.~(\ref{soeq}).
The expressions for these functions are rather cumbersome and their explicit form is presented in Appendix A.

\section{Spin-resolved orbital magnetization}

By analogy to the ordinary magnetization we define the spin-resolved orbital magnetization as the derivative of the free energy $F$ with respect to $B$, $\mathbf{M} = - \partial F/ \partial {\mathbf{B}}$ (see e.g. Refs.~\onlinecite{Niu_2006,ChenLee2011}). Since the magnetic field is a small perturbation, the induced changes in the free energy $F$ and energy $E$ are approximately equal~\cite{LL}, $\delta F \approx \delta E$. Thus, using the  Hellmann-Feynman theorem we can write (see for example Refs.~\onlinecite{resta2005}),
\begin{equation}
\label{Ms}
M_{orb}^{s} = - \frac{\partial \langle H \rangle}{\partial B}.
\end{equation}
In the following we will use the above equation to find the spin-resolved orbital magnetization.

The  quantum-mechanical average of energy, $\langle H \rangle$,  for the system in a spin-dependent magnetic field can be found in the Matsubara-Green's function formalism from the following expression:
\begin{equation}
\label{H_av}
\langle H \rangle = \frac{1}{\beta} \mathrm{Tr} \sum_{n}
\int \frac{d^{2} \mathbf{k}}{(2\pi)^{2}} H_{R} \, \mathcal{G}_{\mathbf{k}}(i\varepsilon_{n}),
\end{equation}
where $\beta = 1/ k_{B} T$ and the Matsubara energies are defined as $ i \varepsilon_{n} = (2n + 1) i \pi k_{B} T$. The sum over Matsubara energies can be calculated by the method of the contour integration~\cite{Machan},
\begin{eqnarray}
\label{Matsubara_sum}
\frac{1}{\beta}\sum_{n}  \hat{H}_{R} \, \mathcal{G}_{\mathbf{k}}(i\varepsilon_{n})
= - \int_{\mathcal{C}} \frac{d z}{2 \pi i} f(z) H_{R} \mathcal{G}_{\mathbf{k}}(z),
\end{eqnarray}
where $f(z)$ is a meromorphic function that has simple poles at the odd integers, $z = i \varepsilon_{n}$, and takes the form $f(z) = ({\mathrm{e}}^{\beta z} + 1)^{-1}$, while $\mathcal{C}$ is the appropriate contour of integration.\cite{Machan}
Combining Eq. (\ref{H_av}) and Eq.(\ref{Matsubara_sum}) one finds
\begin{eqnarray}
\langle H \rangle = -\mathrm{Tr} \int_{\mathcal{C}} \frac{d z}{2\pi i} \int \frac{d^{2} \mathbf{k}}{(2\pi)^{2}} f(z)\,
H_{R} \, \mathcal{G}_{\mathbf{k}}(z).
\end{eqnarray}

The integral along the contour $\mathcal{C}$ has a branch cut at the line $z = \varepsilon$, where $\varepsilon$ is real. Consequently, one can write
\begin{eqnarray}
\langle H \rangle =i\, \mathrm{Tr} \int \frac{d^{2} \mathbf{k}}{(2\pi)^{2}} \int \frac{d \varepsilon}{2\pi} f(\varepsilon)H_{R} \nonumber \\
\times[ \mathcal{G}_{\mathbf{k}}(\varepsilon + i \delta) - \mathcal{G}_{\mathbf{k}}(\varepsilon - i \delta)],
\end{eqnarray}
where $\delta $ is an infinitesimally small positive number. 
After analytical continuation we arrive at the formula
\begin{equation}
\label{Hav}
\langle H \rangle = i \, \mathrm{Tr}\int \frac{d^{2} \mathbf{k}}{(2\pi)^{2}} \int \frac{d \varepsilon}{2\pi}
f(\varepsilon) H_{R}\, [\mathcal{G}^{R}_{\mathbf{k}}(\varepsilon) - \mathcal{G}^{A}_{\mathbf{k}}(\varepsilon)].
\end{equation}
This general expression in combination with the explicit form of the core Green's function (\ref{gf_k}), allows one
to obtain  from  Eq.~(\ref{Ms}) the analytical result for the orbital spin resolved magnetization, which conveniently can be written as a sum of three terms,
\begin{equation}
M_{orb}^{s} = M_{\rm orb}^{s(1)} + M_{\rm orb}^{s(2)} + M_{\rm orb}^{s(3)},
\end{equation}
where
\begin{eqnarray}
\label{Ms1}
M_{\rm orb}^{s(1)} = - 2 i \int \frac{d^{2} \mathbf{k}}{(2\pi)^{2}} \int \frac{d \varepsilon}{2\pi} f(\varepsilon) \varepsilon_{k} \partial_{B}[\mathcal{G}_{\mathbf{k} 0}^{R}(\varepsilon) - \mathcal{G}_{\mathbf{k} 0}^{A}(\varepsilon)],\hspace{0.5cm}\\
\label{Ms2}
M_{\rm orb}^{s(2)} = - 2 i \int \frac{d^{2} \mathbf{k}}{(2\pi)^{2}} \int \frac{d \varepsilon}{2\pi} f(\varepsilon) \alpha k_{y} \partial_{B}[\mathcal{G}_{\mathbf{k} x}^{R}(\varepsilon) - \mathcal{G}_{\mathbf{k} x}^{A}(\varepsilon)],\hspace{0.5cm}\\
\label{Ms3}
M_{\rm orb}^{s(3)} =  2 i \int \frac{d^{2} \mathbf{k}}{(2\pi)^{2}} \int \frac{d \varepsilon}{2\pi} f(\varepsilon) \alpha k_{x} \partial_{B}[\mathcal{G}_{\mathbf{k} y}^{R}(\varepsilon) - \mathcal{G}_{\mathbf{k} y}^{A}(\varepsilon)].\hspace{0.7cm}
\end{eqnarray}
The explicit forms of the integrals in Eqs.~(\ref{Ms1})-(\ref{Ms3}) are given in Appendix B.
After integration over $\varepsilon$ one arrives at (for details see Appendix B)
\begin{eqnarray}
\label{Ms1_B}
M_{\rm orb}^{s(1)} = \alpha \frac{e}{8 \pi \hbar} \int dk \varepsilon_{k}^{2} [f''(E_{+}) - f''(E_{-})]\nonumber\\
- \alpha \frac{e}{8 \pi \hbar} \int dk \frac{\varepsilon_{k}^{2}}{\alpha k} [f'(E_{+}) + f'(E_{-})]\nonumber\\
+ \alpha \frac{e}{8 \pi \hbar} \int dk \frac{\varepsilon_{k}^{2}}{\alpha^{2} k^{2}} [f(E_{+}) - f(E_{-})]\nonumber\\
+ \alpha \frac{e}{8 \pi \hbar} \int dk \frac{\alpha k}{2} \varepsilon_{k} [f''(E_{+}) + f''(E_{-})]\nonumber\\
- \alpha \frac{e}{8 \pi \hbar} \int dk \frac{\varepsilon_{k}}{2} [f'(E_{+}) - f'(E_{-})]
\end{eqnarray}
and
\begin{eqnarray}
\label{Ms23_B}
M_{\rm orb}^{s(2)} + M_{\rm orb}^{s(3)} = \alpha \frac{e}{16 \pi \hbar} \int dk \varepsilon_{k} [f'(E_{+}) - f'(E_{-})]\nonumber\\
+ \alpha \frac{e}{16 \pi \hbar} \int dk \alpha k \varepsilon_{k} [f''(E_{+}) + f''(E_{-})]\nonumber\\
+ \alpha \frac{e}{16 \pi \hbar} \int dk \alpha k [f'(E_{+}) + f'(E_{-})]\nonumber\\
+ \alpha \frac{e}{16 \pi \hbar} \int dk \alpha^{2} k^{2} [f''(E_{+}) - f''(E_{-})]\nonumber\\
- \alpha \frac{e}{16 \pi \hbar} \int dk [f(E_{+}) - f(E_{-})],\hspace{0.5cm}
\end{eqnarray}
where  $f'$ and $f''$ denote the first and second derivatives of the Fermi distribution function with respect to energy.
Upon combining these two equations one finally gets the general expression for the spin-resolved orbital magnetization of the two-dimensional electron gas with Rashba interaction,
\begin{eqnarray}
M_{orb}^{s} = \frac{\alpha e}{16 \pi \hbar} \left[ \int dk \left( \frac{2 \varepsilon_{k}^{2}}{\alpha^{2} k^{2}} - 1\right) [f(E_{+}) - f(E_{-})] \right.\nonumber\\
- \int dk \alpha k \left( \frac{2 \varepsilon_{k}^{2}}{\alpha^{2} k^{2}} - 1\right) [f'(E_{+}) + f'(E_{-})]\nonumber\\
+ \int dk \alpha^{2} k^{2} \left( \frac{2 \varepsilon_{k}^{2}}{\alpha^{2} k^{2}} - 1\right) [f''(E_{+}) - f''(E_{-})] \nonumber\\
+ \left. 2 \int dk \alpha k [E_{+} f''(E_{+}) + E_{-} f''(E_{-})] \right].\hspace{0.5cm}
\end{eqnarray}

Expression (29) is our final result for the orbital spin-resolved magnetization $M_{\rm orb}^{s}$, which is valid at
arbitrary temperature. Though this formula is rather cumbersome, in the zero temperature limit it leads to a simple analytical expression for $M_{orb}^{s} (T = 0) = M_{orb}^{s, T = 0}$. When both subbands are occupied, i.e. when $\mu > 0$, we find the formula (for details see Appendix C)
\begin{equation}
 M_{\rm orb}^{s, T=0} = - \frac{e m \alpha^{2}}{12 \pi \hbar^{3}},
\end{equation}
which means that $M_{\rm orb}^s$ is quadratic in the Rashba parameter $\alpha$.

In Fig.~1 we show the temperature dependence of the orbital magnetization $M^s_{\rm orb}$  normalized to its zero temperature value $M^{s,T=0}_{\rm orb}$.
Different curves correspond to the indicated values of the Fermi energy $\mu_{0}$, i.e. the value of chemical potential at $T=0$. Note that for a fixed  particle density $\rho$, the chemical potential varies with temperature as follows~\cite{Vignale}: $\mu = k_{B} T {\mathrm{ln}}({\mathrm{e}}^{\mu_{0}/k_{B} T} - 1)$ and $\mu_{0} = \pi \hbar^{2} \rho/m$. It is evident that $M^s_{\rm orb}$ diminishes with increasing temperature, and this decrease depends on the Fermi energy (particle density): it is faster for low values of $\mu_{0}$. In turn, variation of the normalized magnetization, $M^s_{\rm orb}/M^{s,T=0}_{\rm orb}$ with increasing Fermi energy $\mu_0$  is shown explicitly in Fig.~2 for several values of temperature. One can observe a saturation of $M^s_{\rm orb}$ at its low temperature value when the  particle density is sufficiently large.
\begin{figure}[]
     \centering
     \includegraphics[width=0.9\columnwidth]{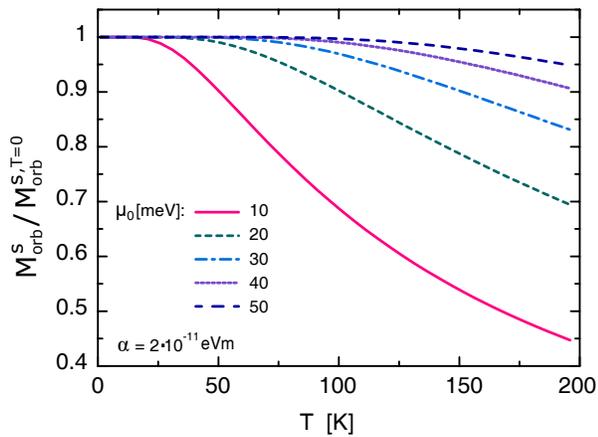}
     \caption{Spin-resolved orbital magnetization $M^{s}_{\rm orb}$,
normalized to its zero-temperature value $M^{s, T=0}_{\rm orb}$, plotted as
a function of temperature for fixed  values of the Fermi level, $\mu_{0}$, as
indicated. Other parameters are: $m = 0.07\, m_{0}$ (where $m_{0}$ is the
electron mass), and $\alpha = 2 \times 10^{-11}$ eV m.}
     \label{fig:fig1}
\end{figure}
\begin{figure}[t]
     \centering
     \includegraphics[width=0.9\columnwidth]{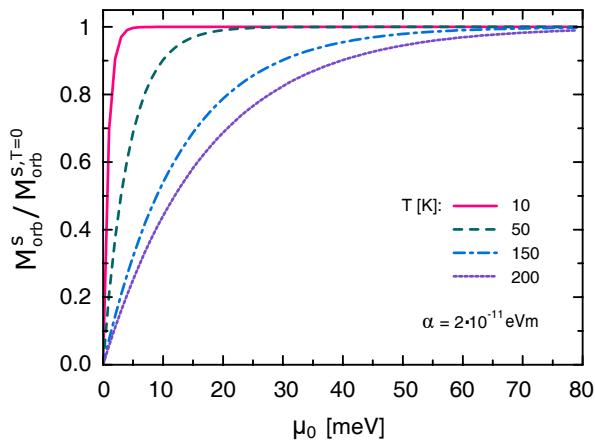}
     \caption{Spin-resolved orbital magnetization $M^{s}_{orb}$,
normalized to its zero-temperature value $M^{s, T=0}_{orb}$, plotted as
a function of the Fermi level $\mu_{0}$ and for fixed temperatures, as
indicated. The other parameters as in Fig.~1.}
     \label{fig:fig2}
\end{figure}

The physical reason for the appearance of spin-resolved orbital magnetization is
related to non-compensated spin currents flowing at the edge of a sample.
Note, that when the temperature is homogeneous, the spin currents are compensated in the bulk except the edges.
However,  when the temperature is non-homogeneous, the spin currents can also exist in the bulk
(see the discussions below).
This is a spin analogy to the usual orbital magnetization, which arises due to  non-compensated electric currents
at the edges.

\section{Discussion and conclusions}

In our recent paper~\cite{Dyrdal2016}  we used the Matsubara Green's function method to calculate the spin Nernst conductivity $\alpha^{s_z}_{xy}$.
This conductivity defines spin current flowing perpendicularly to the temperature gradient. We have shown there that the vertex correction due to scattering on impurities does not cancel the {\it bare bubble} contribution, contrary to the spin Hall conductivity where such a cancelation takes place.
As a result, the spin Nernst conductivity in this approximation diverges in the zero-temperature limit.

To remove this divergency it
was necessary to include an additional contribution to the spin current (and also to the spin Nernst conductivity) that follows from the spin-resolved orbital magnetization. One can conceive the spin current ${\bf J}^z$ as a superposition of spin-up and spin-down currents
flowing in opposite directions, ${\bf J}^z={\bf J}_\uparrow -{\bf J}_\downarrow $.
Each of the spin-polarized currents generates the corresponding orbital magnetization.
However, the vectors ${\bf M}_\uparrow $ and ${\bf M}_\downarrow $ are oriented in opposite directions, so the
total orbital magnetization vanishes, ${\bf M}={\bf M}_\uparrow +{\bf M}_\downarrow=0$, as one can expect from the
time-reversion symmetry. In turn, the spin-resolved orbital magnetization,
${\bf M}_{orb}^{s}={\bf M}_\uparrow -{\bf M}_\downarrow $, is nonzero, ${\bf M}_{orb}^{s}\ne 0$.

The spin current due to spin-resolved orbital magnetization depends on temperature. Therefore, it contributes to spin Nernst conductivity as the
corresponding currents flowing at the edges having different temperatures do not cancel each other, though they flow in opposite directions.
In turn, these currents do not contribute to the spin Hall conductivity because in thermally uniform system the current at the two edges cancel each other. The correction to the spin Nernst conductivity that originates from the spin-resolved orbital magnetization is given by the term $(\hbar/e) M_{orb}^{s}/T$

Similar situation takes place also in the case of Nernst effect in systems with no time-reversal symmetry. In that case the absence of time-reversal symmetry
admits orbital magnetization. This magnetization in turn contributes to charge current, and the corresponding contribution removes the zero-temperature divergency in the Nernst conductivity (see e.g. Refs.~\onlinecite{obraztsov,Streda_1985,Varlamov2014,Varlamov2015})

\begin{acknowledgments}
This work was supported by the National Science Center in Poland as the Project
No. DEC-2012/04/A/ST3/00372, and partly   by the Polish Ministry of Science and Higher Education through a research project ’Iuventus
Plus’ in years 2015-2017 (project No. 0083/IP3/2015/73). A.D. acknowledges the support from the Fundation for Polish Science (FNP).
\end{acknowledgments}

\appendix
\onecolumngrid

\section{Equations for the Green Function and its Fourier Transform}

In this appendix we derive the matrix equation (10). Multiplying Eq.(8) on the right by ${\rm e}^{-i \sigma_{z} \mathcal{A}_{\mathbf{r} \mathbf{r''}}}$ and using commutation relations for the Pauli matrices we find
\begin{eqnarray}
 {\rm e}^{i \sigma_{z} \mathcal{A}_{\mathbf{r} \mathbf{r}'}}\mathcal{G}_{0}(\varepsilon, \mathbf{r}', \mathbf{r}''){\rm e}^{i \sigma_{z} \mathcal{A}_{\mathbf{r'} \mathbf{r''}}} {\rm e}^{-i \sigma_{z} \mathcal{A}_{\mathbf{r} \mathbf{r''}}} = \mathcal{G}_{0 0}(\varepsilon, \mathbf{r}', \mathbf{r}'') \sigma_{0} \mathrm{e}^{i \sigma_{z} (\mathcal{A}_{\mathbf{r} \mathbf{r}'} + \mathcal{A}_{\mathbf{r}' \mathbf{r}'} -\mathcal{A}_{\mathbf{r} \mathbf{r}''} )} + \mathcal{G}_{0 z}(\varepsilon, \mathbf{r}', \mathbf{r}'') \sigma_{z} \mathrm{e}^{i \sigma_{z} (\mathcal{A}_{\mathbf{r} \mathbf{r}'} + \mathcal{A}_{\mathbf{r}' \mathbf{r}'} -\mathcal{A}_{\mathbf{r} \mathbf{r}''} )} \nonumber\\
+ \mathcal{G}_{0 x}(\varepsilon, \mathbf{r}', \mathbf{r}'') \sigma_{x} \mathrm{e}^{-i \sigma_{z} (\mathcal{A}_{\mathbf{r} \mathbf{r}'} - \mathcal{A}_{\mathbf{r}' \mathbf{r}''} - \mathcal{A}_{\mathbf{r} \mathbf{r}''})} + \mathcal{G}_{0 y}(\varepsilon, \mathbf{r}', \mathbf{r}'') \sigma_{y} \mathrm{e}^{-i \sigma_{z} (\mathcal{A}_{\mathbf{r} \mathbf{r}'} - \mathcal{A}_{\mathbf{r}' \mathbf{r}''} - \mathcal{A}_{\mathbf{r} \mathbf{r}''})}.
\end{eqnarray}

The integral along the contour  $\mathbf{r}-\mathbf{r}'-\mathbf{r}''-\mathbf{r}$ in the Peierls phase can be transformed into the surface integral,
\begin{equation}
 \frac{e}{\hbar}\oint\mathbf{A}({\mathbf{R}})\cdot d{\mathbf{R}} 
 = \frac{e}{\hbar} \left( \int_{\mathbf{r}}^{\mathbf{r}'} \mathbf{A}({\mathbf{R}})\cdot d{\mathbf{R}} + \int_{\mathbf{r}'}^{\mathbf{r''}} \mathbf{A}({\mathbf{R}})\cdot d{\mathbf{R}} + \int_{\mathbf{r}''}^{\mathbf{r}} \mathbf{A}({\mathbf{R}})\cdot d{\mathbf{R}}\right) = \frac{e}{\hbar} \frac{1}{2} \mathbf{B} \cdot (\mathbf{r}' - \mathbf{r}) \times (\mathbf{r}'' - \mathbf{r}'),
\end{equation}
so Eq.~(A1) takes the form
\begin{eqnarray}
 {\rm e}^{i \sigma_{z} \mathcal{A}_{\mathbf{r} \mathbf{r}'}}\mathcal{G}_{0}(\varepsilon, \mathbf{r}', \mathbf{r}''){\rm e}^{i \sigma_{z} \mathcal{A}_{\mathbf{r'} \mathbf{r''}}} {\rm e}^{-i \sigma_{z} \mathcal{A}_{\mathbf{r} \mathbf{r''}}} = \mathcal{G}_{0 0}(\varepsilon, \mathbf{r}', \mathbf{r}'') \sigma_{0} {\mathrm{e}}^{i \sigma_{z} \frac{e}{\hbar}  \mathbf{B} \cdot \frac{1}{2} (\mathbf{r}' - \mathbf{r})\times(\mathbf{r}'' - \mathbf{r}') }
+ \mathcal{G}_{0 x}(\varepsilon, \mathbf{r}', \mathbf{r}'') \sigma_{x} + \mathcal{G}_{0 y}(\varepsilon, \mathbf{r}', \mathbf{r}'') \sigma_{y} \nonumber\\
+ \mathcal{G}_{0 z}(\varepsilon, \mathbf{r}', \mathbf{r}'') \sigma_{z} {\mathrm{e}}^{i \sigma_{z} \frac{e}{\hbar}  \mathbf{B} \cdot \frac{1}{2} (\mathbf{r}' - \mathbf{r})\times(\mathbf{r}'' - \mathbf{r}') }.
\hskip0.5cm
\end{eqnarray}

Inserting Eq.(A2)  into Eq.~(8) we obtain a
set of four equations for the four components of the core Green function $\mathcal{G}_{0 i}(\varepsilon, \mathbf{r}',\mathbf{r}'')$,
\begin{subequations}
\begin{align}
\int d \mathbf{r}' (\varepsilon - H_{0}) \mathcal{G}_{0 0}(\varepsilon, \mathbf{r}', \mathbf{r}'') \delta(\mathbf{r} - \mathbf{r}') + i \int d \mathbf{r}'(\varepsilon - H_{0})\mathcal{G}_{0 z}(\varepsilon, \mathbf{r}', \mathbf{r}'') \frac{e}{2 \hbar} \mathbf{B} \cdot   (\mathbf{r}' - \mathbf{r}) \times (\mathbf{r}'' - \mathbf{r}') \delta(\mathbf{r} - \mathbf{r}')\nonumber\\
 - \alpha \int d \mathbf{r}' \kappa_{y}  \mathcal{G}_{0 x}(\varepsilon, \mathbf{r}', \mathbf{r}'') \delta(\mathbf{r} - \mathbf{r}') + \alpha \int d \mathbf{r}' \kappa_{x}  \mathcal{G}_{0 y}(\varepsilon, \mathbf{r}', \mathbf{r}'') \delta(\mathbf{r} - \mathbf{r}') = \delta(\mathbf{r} - \mathbf{r}''),
\end{align}
\begin{align}
- \alpha \int d \mathbf{r}'  \kappa_{y} \mathcal{G}_{0 0}(\varepsilon, \mathbf{r}', \mathbf{r}'') \delta(\mathbf{r} - \mathbf{r}') - \alpha \int d \mathbf{r}' \kappa_{x}  \mathcal{G}_{0 0}(\varepsilon, \mathbf{r}', \mathbf{r}'') \frac{e}{2 \hbar} \mathbf{B} \cdot (\mathbf{r}' - \mathbf{r}) \times (\mathbf{r}'' - \mathbf{r}') \delta(\mathbf{r} - \mathbf{r}')\nonumber\\
- i \alpha \int d \mathbf{r}' \kappa_{y} \mathcal{G}_{0 z}(\varepsilon, \mathbf{r}', \mathbf{r}'') \frac{e}{2 \hbar}  \mathbf{B} \cdot (\mathbf{r}' - \mathbf{r}) \times (\mathbf{r}'' - \mathbf{r}') \delta(\mathbf{r} - \mathbf{r}') + i \alpha \int d \mathbf{r}' \kappa_{x}  \mathcal{G}_{0 z}(\varepsilon, \mathbf{r}', \mathbf{r}'') \delta(\mathbf{r} - \mathbf{r}')\nonumber\\
+\int d \mathbf{r}' (\varepsilon - H_{0}) \mathcal{G}_{0 x}(\varepsilon, \mathbf{r}', \mathbf{r}'') \delta(\mathbf{r} - \mathbf{r}') = 0,
\end{align}
\begin{align}
- \alpha \int d \mathbf{r}' \kappa_{y}  \mathcal{G}_{0 0}(\varepsilon, \mathbf{r}', \mathbf{r}'') \frac{e}{2 \hbar}  \mathbf{B} \cdot (\mathbf{r}' - \mathbf{r}) \times (\mathbf{r}'' - \mathbf{r}') \delta(\mathbf{r} - \mathbf{r}') + \alpha \int d \mathbf{r}' \kappa_{x} \mathcal{G}_{0 0}(\varepsilon, \mathbf{r}', \mathbf{r}'') \delta(\mathbf{r} - \mathbf{r}') \nonumber\\
+ i \alpha \int d \mathbf{r}' \kappa_{y} \mathcal{G}_{0 z}(\varepsilon, \mathbf{r}', \mathbf{r}'') \delta(\mathbf{r} - \mathbf{r}') + i \alpha \int d \mathbf{r}' \kappa_{x} \mathcal{G}_{0 z}(\varepsilon, \mathbf{r}', \mathbf{r}'') \frac{e}{2 \hbar} \mathbf{B} \cdot (\mathbf{r}' - \mathbf{r})\times (\mathbf{r}'' - \mathbf{r}') \delta(\mathbf{r} - \mathbf{r}') \nonumber\\
+ \int d \mathbf{r}' (\varepsilon - H_{0}) \mathcal{G}_{0 y}(\varepsilon, \mathbf{r}', \mathbf{r}'')\delta(\mathbf{r} - \mathbf{r}') = 0,
\end{align}
\begin{align}
i \int d \mathbf{r}' (\varepsilon - H_{0}) \mathcal{G}_{0 0}(\varepsilon, \mathbf{r}', \mathbf{r}'') \frac{e}{2 \hbar} \mathbf{B} \cdot (\mathbf{r}' - \mathbf{r}) \times (\mathbf{r}'' - \mathbf{r}') \delta(\mathbf{r} - \mathbf{r}') + \int d \mathbf{r}' (\varepsilon - H_{0}) \mathcal{G}_{0 z}(\varepsilon, \mathbf{r}', \mathbf{r}'') \delta(\mathbf{r} - \mathbf{r}') \nonumber\\
- i \alpha \int d \mathbf{r}' \kappa_{x} \mathcal{G}_{0 x}(\varepsilon, \mathbf{r}', \mathbf{r}'')\delta(\mathbf{r} - \mathbf{r}') - i \alpha  \int d \mathbf{r}'  \kappa_{y} \mathcal{G}_{0 y}(\varepsilon, \mathbf{r}', \mathbf{r}'')\delta(\mathbf{r} - \mathbf{r}') = 0.
\end{align}
\end{subequations}
where  $H_{0} = \hbar^{2} (\kappa_{x}^{2} + \kappa_{y}^{2})/2m$,  with $\kappa_{\alpha} = - i \nabla_{\alpha}$ , and we expanded the exponential factors to the first order in $B$.

After  Fourier transformation, this set of equations takes the form
{\small{
\begin{subequations}
\begin{align}
[g_{\mathbf{k} 0}(\varepsilon)]^{-1} \mathcal{G}_{\mathbf{k} 0}(\varepsilon) - i \frac{e}{2 \hbar} B_{k} \epsilon_{ijk}\left(\frac{\partial }{\partial k_{i}} [g_{\mathbf{k} 0}(\varepsilon)]^{-1} \right) \left( \frac{\partial }{\partial k_{j}} \mathcal{G}_{\mathbf{k} z} (\varepsilon)\right) - \alpha k_{y} \mathcal{G}_{\mathbf{k} x}(\varepsilon) + \alpha k_{x} \mathcal{G}_{\mathbf{k} y}(\varepsilon) = 1,
\end{align}
\begin{align}
- \alpha k_{y} \mathcal{G}_{\mathbf{k} 0}(\varepsilon) + \alpha \frac{e}{2 \hbar} B_{k} \epsilon_{ijk} \delta_{ix} \left( \frac{\partial}{\partial k_{j}} \mathcal{G}_{\mathbf{k} 0}(\varepsilon)\right) + i \alpha \frac{e}{2 \hbar}  B_{k} \epsilon_{ijk} \delta_{i y} \left( \frac{\partial}{\partial k_{j}} \mathcal{G}_{\mathbf{k} z}(\varepsilon)\right) + i \alpha k_{x} \mathcal{G}_{\mathbf{k} z}(\varepsilon \mathbf{k}) + [g_{\mathbf{k} 0}(\varepsilon)]^{-1} \mathcal{G}_{\mathbf{k} x}(\varepsilon, \mathbf{k}) = 0,
\end{align}
\begin{align}
\alpha \frac{e}{2\hbar}  B_{k} \epsilon_{ijk} \delta_{i y} \left( \frac{\partial}{\partial k_{j}} \mathcal{G}_{\mathbf{k} 0}(\varepsilon)\right) + \alpha k_{x} \mathcal{G}_{\mathbf{k} 0}(\varepsilon) + i \alpha k_{y} \mathcal{G}_{\mathbf{k} z}(\varepsilon) - i \alpha \frac{e}{2 \hbar}  B_{k} \epsilon_{ijk} \delta_{i x} \left( \frac{\partial}{\partial k_{j}} \mathcal{G}_{\mathbf{k} z}(\varepsilon)\right) +[g_{\mathbf{k} 0}(\varepsilon)]^{-1} \mathcal{G}_{\mathbf{k} y}(\varepsilon) = 0,
\end{align}
\begin{align}
-i \frac{e}{2 \hbar} B_{k} \epsilon_{ijk} \left( \frac{\partial}{\partial k_{i}} [g_{\mathbf{k} 0}(\varepsilon)]^{-1}\right) \left( \frac{\partial}{\partial k_{j}} \mathcal{G}_{\mathbf{k} 0}(\varepsilon)\right) + [g_{\mathbf{k} 0}(\varepsilon)]^{-1} \mathcal{G}_{\mathbf{k} z}(\varepsilon) - i \alpha k_{x} \mathcal{G}_{\mathbf{k} x}(\varepsilon) - i \alpha k_{y} \mathcal{G}_{\mathbf{k} y}(\varepsilon) = 0,
\end{align}
\end{subequations}
}}
where $[g_{\mathbf{k} 0}(\varepsilon)]^{-1} = \varepsilon - \varepsilon_{k}$.

Equations (A4a)-(A4d) may be further simplified assuming linear response with respect to $B$,
\begin{subequations}
\begin{align}
 i \frac{e}{2 \hbar}  B_{k}  \epsilon_{ijk}  \left( \frac{\partial}{\partial k_{i}} [g_{\mathbf{k} 0}(\varepsilon)]^{-1}\right)\left( \frac{\partial}{\partial k_{j}} \mathcal{G}_{\mathbf{k} z}(\varepsilon)\right) \cong  i \frac{e}{2 \hbar} B \epsilon_{ijk}  \left( \frac{\partial}{\partial k_{i}} [g_{\mathbf{k} 0}(\varepsilon)]^{-1}\right) \left( \frac{\partial}{\partial k_{j}} G_{\mathbf{k} z}(\varepsilon)\right) = 0,
\end{align}
\begin{align}
\alpha \frac{e}{2 \hbar}  B_{k} \epsilon_{ijk} \delta_{i x,y} \left( \frac{\partial}{\partial k_{j}} \mathcal{G}_{\mathbf{k} 0}(\varepsilon)\right) \cong \alpha \frac{e}{2 \hbar}  B_{k}  \epsilon_{ijk} \delta_{i x,y} \left( \frac{\partial}{\partial k_{j}} G_{\mathbf{k} 0}(\varepsilon)\right) ,\hspace{5.5cm}
\end{align}
\begin{align}
i \alpha \frac{e}{2 \hbar}  B_{k}  \epsilon_{ijk} \delta_{i x,y} \left( \frac{\partial}{\partial k_{j}} \mathcal{G}_{\mathbf{k} z}\right) \cong i \alpha \frac{e}{2 \hbar}  B_{k}  \epsilon_{ijk} \delta_{i x,y} \left( \frac{\partial}{\partial k_{j}} G_{\mathbf{k} z}\right) = 0,\hspace{6.2cm}
\end{align}
\begin{align}
i \frac{e}{2 \hbar}  B_{k}  \epsilon_{ijk} \left( \frac{\partial}{\partial k_{i}} [g_{\mathbf{k} 0}(\varepsilon)]^{-1}\right) \left( \frac{\partial}{\partial k_{j}} \mathcal{G}_{\mathbf{k} 0}(\varepsilon)\right) \cong i \frac{e}{2 \hbar}  B_{k}  \epsilon_{ijk} \left( \frac{\partial}{\partial k_{i}} [g_{\mathbf{k} 0}(\varepsilon)]^{-1}\right) \left( \frac{\partial}{\partial k_{j}} G_{\mathbf{k} 0}(\varepsilon)\right),\hspace{1.3cm}
\end{align}
\end{subequations}
and finally we obtain the matrix equation (\ref{soeq}).

\section{Integration over $\varepsilon$}
%
As follows from  Eqs.~(27) and (28), we need to calculate 24 integrals over $\varepsilon$:
\begin{eqnarray}
\mathcal{I}_{1,2} = \int d\varepsilon \frac{f(\varepsilon)}{\varepsilon + \mu - E_{\pm} + i \Gamma} = \mathcal{P}\int d\varepsilon \frac{f(\varepsilon)}{\varepsilon + \mu - E_{\pm}} - i \pi f(E_{\pm}),\hspace{8.8cm}\\ \nonumber\\
\mathcal{I}_{3,4} = \int d\varepsilon \frac{f(\varepsilon)}{\varepsilon + \mu - E_{\pm} - i \Gamma} = \mathcal{P}\int d\varepsilon \frac{f(\varepsilon)}{\varepsilon + \mu - E_{\pm}} + i \pi f(E_{\pm}),\hspace{8.8cm}\\ \nonumber\\
\mathcal{I}_{5,6} = \int d\varepsilon \frac{f(\varepsilon)}{(\varepsilon + \mu - E_{\pm} + i \Gamma)^{2}} = \mathcal{P} \int d\varepsilon \frac{\partial f(\varepsilon)}{\partial \varepsilon} \frac{1}{\varepsilon + \mu - E_{\pm}} - i \pi f'(E_{\pm}),\hspace{7.3cm}\\ \nonumber\\
\mathcal{I}_{7,8} = \int d\varepsilon \frac{f(\varepsilon)}{(\varepsilon + \mu - E_{\pm} - i \Gamma)^{2}} = \mathcal{P} \int d\varepsilon \frac{\partial f(\varepsilon)}{\partial \varepsilon} \frac{1}{\varepsilon + \mu - E_{\pm}} + i \pi f'(E_{\pm}),\hspace{7.3cm}\\ \nonumber\\
\mathcal{I}_{9,10} = \int d\varepsilon \frac{f(\varepsilon)}{(\varepsilon + \mu - E_{\pm} + i \Gamma)^{3}} = \frac{1}{2} \mathcal{P} \int d\varepsilon \frac{\partial^{2} f(\varepsilon)}{\partial \varepsilon^{2}} \frac{1}{\varepsilon + \mu - E_{\pm}} - i \frac{\pi}{2} f''(E_{\pm}),\hspace{6.5cm}\\ \nonumber\\
\mathcal{I}_{11,12} = \int d\varepsilon \frac{f(\varepsilon)}{(\varepsilon + \mu - E_{\pm} - i \Gamma)^{3}} = \frac{1}{2} \mathcal{P} \int d\varepsilon \frac{\partial^{2} f(\varepsilon)}{\partial \varepsilon^{2}} \frac{1}{\varepsilon + \mu - E_{\pm}} + i \frac{\pi}{2} f''(E_{\pm}),\hspace{6.5cm}\\ \nonumber\\
\mathcal{I}_{13,14} = \int d\varepsilon \frac{(\varepsilon + \mu - \varepsilon_{k})f(\varepsilon)}{(\varepsilon + \mu - E_{\pm} + i \Gamma)^{3}} = \frac{1}{2}
\mathcal{P} \int d\varepsilon [2 f'(\varepsilon) + (\varepsilon + \mu - \varepsilon_{k}) f''(\varepsilon)]\frac{1}{\varepsilon + \mu - E_{\pm}} \mp i \frac{\pi}{2}[2 f'(E_{\pm})  + \alpha k  f''(E_{\pm})],\hspace{1cm}\\ \nonumber\\
\mathcal{I}_{15,16} = \int d\varepsilon \frac{(\varepsilon + \mu - \varepsilon_{k})f(\varepsilon)}{(\varepsilon + \mu - E_{-} - i \Gamma)^{3}} = \frac{1}{2} \mathcal{P} \int d\varepsilon [2 f'(\varepsilon) + (\varepsilon + \mu - \varepsilon_{k}) f''(\varepsilon)]\frac{1}{\varepsilon + \mu - E_{-}} \mp i \frac{\pi}{2}[2 f'(E_{-})  - \alpha k  f''(E_{-})],\hspace{1cm}\\ \nonumber\\
\mathcal{I}_{17,18} = \int d\varepsilon \frac{(\varepsilon + \mu - \varepsilon_{k}) f(\varepsilon)}{(\varepsilon + \mu - E_{+} \pm i\Gamma)^{2}} = \mathcal{P} \int d\varepsilon [f(\varepsilon) + (\varepsilon + \mu - \varepsilon_{k}) f'(\varepsilon)] \frac{1}{\varepsilon + \mu - E_{+}} \mp i \pi  [f(E_{+}) + \alpha k f'(E_{+})],\hspace{2cm}\\ \nonumber\\
\mathcal{I}_{19, 20} = \int d\varepsilon \frac{(\varepsilon + \mu - \varepsilon_{k}) f(\varepsilon)}{(\varepsilon + \mu - E_{-} \pm i \Gamma)^{2}} = \mathcal{P} \int d\varepsilon [f(\varepsilon) + (\varepsilon + \mu - \varepsilon_{k}) f'(\varepsilon)] \frac{1}{\varepsilon + \mu - E_{-}} \mp i \pi [f(E_{-}) - \alpha k f'(E_{-})],\hspace{2cm}\\ \nonumber\\
\mathcal{I}_{21,22} = \int d\varepsilon \frac{f(\varepsilon) (\varepsilon + \mu - \varepsilon_{k})}{\varepsilon + \mu - E_{\pm} + i \Gamma} = \mathcal{P} \int d\varepsilon \frac{f(\varepsilon) (\varepsilon + \mu - \varepsilon_{k})}{\varepsilon + \mu - E_{\pm}} \mp i \pi \alpha k f(E_{\pm}),\hspace{7.3cm}\\ \nonumber\\
\mathcal{I}_{23,24} = \int d\varepsilon \frac{f(\varepsilon) (\varepsilon + \mu - \varepsilon_{k})}{\varepsilon + \mu - E_{\pm} - i \Gamma} = \mathcal{P} \int d\varepsilon \frac{f(\varepsilon) (\varepsilon + \mu - \varepsilon_{k})}{\varepsilon + \mu - E_{\pm}} \pm i \pi \alpha k f(E_{\pm}),\hspace{7.3cm}
\end{eqnarray}

According to the above,  we may write $M_{\rm orb}^{s(1)}$ and $M_{\rm orb}^{s(2)} + M_{\rm orb}^{s(3)}$ as follows:
\begin{eqnarray}
M_{\rm orb}^{s(1)} =  i \frac{\alpha e}{4 \pi \hbar} \int \frac{d k}{2\pi} \varepsilon_{k}^{2} [\mathcal{I}_{9} - \mathcal{I}_{11} + \mathcal{I}_{12} - \mathcal{I}_{10} + \frac{1}{2 \alpha k} (-\mathcal{I}_{5} + \mathcal{I}_{7} - \mathcal{I}_{6} + \mathcal{I}_{8}) + \frac{2}{(2\alpha k)^{2}} (\mathcal{I}_{1} - \mathcal{I}_{3} - \mathcal{I}_{2} + \mathcal{I}_{4})] \nonumber\\
+ i \frac{\alpha e}{4 \pi \hbar} \int \frac{d k}{2\pi} \frac{\alpha k}{2} \varepsilon_{k} [\mathcal{I}_{9} - \mathcal{I}_{11} - \mathcal{I}_{12} + \mathcal{I}_{10} + \frac{1}{2\alpha k}(- \mathcal{I}_{5} + \mathcal{I}_{7} + \mathcal{I}_{6} - \mathcal{I}_{8})],
\end{eqnarray}
\begin{eqnarray}
M_{\rm orb}^{s(2)} + M_{\rm orb}^{s(3)} = i \frac{\alpha e}{8 \pi \hbar} \int \frac{dk}{2\pi} E_{+} [\mathcal{I}_{13} - \mathcal{I}_{14} + \frac{1}{4 \alpha^{2} k^{2}} (\mathcal{I}_{21} - \mathcal{I}_{22} - \mathcal{I}_{23} + \mathcal{I}_{24}) - \frac{1}{2 \alpha k}(\mathcal{I}_{17} - \mathcal{I}_{18})]\nonumber\\
+ i \frac{\alpha e}{8 \pi \hbar} \int \frac{dk}{2\pi} E_{-} [\mathcal{I}_{16} - \mathcal{I}_{15} - \frac{1}{4 \alpha^{2} k^{2}} (\mathcal{I}_{22} - \mathcal{I}_{21} - \mathcal{I}_{24} + \mathcal{I}_{23}) - \frac{1}{2\alpha k} (\mathcal{I}_{19} - \mathcal{I}_{20})].
\end{eqnarray}
Taking into account explicit forms of the integrals $\mathcal{I}_{n}$ we find:
\begin{eqnarray}
M_{\rm orb}^{s(1)} =  i \frac{\alpha e}{4 \pi \hbar} \int \frac{d k}{2\pi} \varepsilon_{k}^{2} [\pi(f''(E_{+}) - f''(E_{-})) - \frac{\pi}{\alpha k} (f'(E_{+}) + f'(E_{-})) + \frac{\pi}{\alpha^{2} k^{2}}(f(E_{+}) - f(E_{-}))] \nonumber\\
+  \frac{\alpha e}{4 \pi \hbar} \int \frac{d k}{2\pi} \frac{\alpha k}{2} \varepsilon_{k}[\pi (f''(E_{+}) + f''(E_{-})) - \frac{\pi}{\alpha k} (f'(E_{+}) - f'(E_{-}))],
\end{eqnarray}
\begin{eqnarray}
M_{\rm orb}^{s(2)} + M_{\rm orb}^{s(3)} =  \frac{\alpha e}{8 \pi \hbar} \int \frac{dk}{2\pi} E_{+} [\pi (2 f'(E_{+}) + \alpha k f''(E_{+})) + \frac{\pi}{2 \alpha k} (f(E_{+}) + f(E_{-})) - \frac{\pi}{\alpha k} (f(E_{+}) + \alpha k f'(E_{+}))]\nonumber\\
+  \frac{\alpha e}{8 \pi \hbar} \int \frac{dk}{2\pi} E_{-}[- \pi (2 f'(E_{-}) - \alpha k f''(E_{-})) + \frac{\pi}{2 \alpha k} (f(E_{-}) + f(E_{+})) - \frac{\pi}{\alpha k} (f(E_{-}) - \alpha k f'(E_{-}))] .\hspace{0.5cm}
\end{eqnarray}
From Eqs. (B15) and (B16) one finally arrives at Eqs.~(\ref{Ms1_B}) and (\ref{Ms23_B}), respectively.

\section{Spin-resolved orbital magnetization in the zero-temperature limit}

In the low temperature limit the orbital magnetization $M_{orb}^{s}$ can be calculated analytically. To do this let us write
$M_{\rm orb}^{s}$ in the form (see Eqs~(27) and (28),
\begin{equation}
M_{\rm orb}^{s(1)} = \sum_{i=1}^{5} \mathcal{M}_{i}\, ,
\end{equation}
\begin{equation}
M_{\rm orb}^{s(2)} + M_{\rm orb}^{s(3)} = \sum_{i=6}^{10} \mathcal{M}_{i}\, ,
\end{equation}
where
\begin{eqnarray}
\mathcal{M}_{1} = \frac{e \alpha}{8 \pi \hbar} \int dk \varepsilon_{k}^{2} [f''(E_{+}) - f''(E_{-})]\hspace{10cm}\nonumber\\
= \frac{e \hbar^{3} \alpha}{32 \pi m \sqrt{m^{2} \alpha^{2} + 2 m \mu \hbar^{2}}} \left[ \int dk \frac{\partial}{\partial k}\left(k^{4} \frac{\partial k}{\partial E_{+}}\right) \delta(k - k_{+}) - \int dk \frac{\partial}{\partial k} \left(k^{4} \frac{\partial k}{\partial E_{-}}\right) \delta(k - k_{-}) \right]
= - \frac{2 m \alpha^{2}}{4\pi\hbar^{3}},
\end{eqnarray}
\begin{eqnarray}
\mathcal{M}_{2} = - \frac{e \hbar^{3}}{32 \pi m^{2}} \int dk k^{3}[f'(E_{+}) + f'(E_{-})]
= \frac{e \hbar^{3}}{32 \pi m \sqrt{m^{2} \alpha^{2} + 2 m \mu \hbar^{2}}} \int dk k^{3} [\delta(k - k_{+}) + \delta(k - k_{-})]\hspace{1cm}\nonumber\\
= \frac{e m \alpha^{2}}{4 \pi \hbar^{3}} + \frac{e \mu}{8 \pi \hbar},
\end{eqnarray}
\begin{eqnarray}
\mathcal{M}_{3} = \alpha \frac{e}{8 \pi \hbar} \int dk \frac{\varepsilon_{k}^{2}}{\alpha^{2} k^{2}} [f(E_{+}) - f(E_{-})]
= \frac{e \hbar^{3}}{32 \pi m^{2} \alpha} \int_{k_{-}}^{k_{+}} dk k^{2} = - \frac{e m \alpha^{2}}{12 \pi \hbar^{3}} - \frac{e \mu}{8 \pi \hbar},\hspace{3.6cm}
\end{eqnarray}
\begin{eqnarray}
\mathcal{M}_{4} = \frac{e \alpha^{2}}{16 \pi \hbar} \int dk k \varepsilon_{k} [f''(E_{+}) + f''(E_{-})] \hspace{9.7cm}
\nonumber\\
= \frac{e \alpha^{2} \hbar}{32 \pi \sqrt{m^{2} \alpha^{2} + 2 m \mu \hbar^{2}}} \left[ \int dk \frac{\partial}{\partial k} \left( k^{3} \frac{\partial k}{\partial E_{+}}\right) \delta(k - k_{+}) + \int dk \frac{\partial}{\partial k} \left( k^{3} \frac{\partial k}{\partial E_{-}}\right) \delta(k - k_{-})\right]
= \frac{e \alpha^{2} m}{8 \pi \hbar^{3}},
\end{eqnarray}
\begin{eqnarray}
\mathcal{M}_{5} = - \frac{e \alpha}{16 \pi \hbar} \int dk \varepsilon_{k} [f'(E_{+}) - f'(E_{-})]
= \frac{e \alpha \hbar}{32 \pi \sqrt{m^{2} \alpha^{2} + 2 m \mu \hbar^{2}}} \int dk k^{2} [\delta(k - k_{+}) - \delta(k - k_{-})]
= - \frac{e m \alpha^{2}}{8 \pi \hbar^{3}},
\end{eqnarray}
\begin{eqnarray}
\mathcal{M}_{6} = - \mathcal{M}_{5},\hspace{14,5cm}
\end{eqnarray} 
\begin{eqnarray}
\mathcal{M}_{7} = \mathcal{M}_{4},\hspace{15cm}
\end{eqnarray}
\begin{eqnarray}
\mathcal{M}_{8} = \frac{e \alpha}{16 \pi \hbar} \int dk \alpha k [f'(E_{+}) + f'(E_{-})]\hspace{9cm}\nonumber\\
= - \frac{e \alpha^{2} m}{16 \pi \hbar \sqrt{m^{2} \alpha^{2} + 2 m \mu \hbar^{2}}} \int dk k [\delta(k - k_{+}) + \delta(k - k_{-})]
= - \frac{m \alpha^{2} e}{8 \pi \hbar^{3}},
\end{eqnarray}
\begin{eqnarray}
\mathcal{M}_{9} = \frac{e \alpha}{16 \pi \hbar}\int dk \alpha^{2} k^{2} [f''(E_{+}) - f''(E_{-})]\hspace{9cm}\nonumber\\
= \frac{\alpha^{3} m e}{16 \pi \hbar \sqrt{m^{2} \alpha^{2} + 2 m \mu \hbar^{2}}} \left[ \int dk \frac{\partial}{\partial k} \left(k^{2} \frac{\partial k}{\partial E_{+}} \right) \delta(k - k_{+}) - \int dk \frac{\partial}{\partial k} \left( k^{2} \frac{\partial k}{\partial E_{-}}\right) \delta(k - k_{-}) \right]
= 0,
\end{eqnarray}
\begin{eqnarray}
\mathcal{M}_{10} = - \frac{e \alpha}{16 \pi \hbar} \int dk [f(E_{+}) - f(E_{-})]
= - \frac{e \alpha}{16 \pi \hbar} \int_{k_{-}}^{k_{+}} dk = \frac{\alpha^{2} m e}{8 \pi \hbar^{3}}.\hspace{5.5cm}
\end{eqnarray}
Taking into account the above results one arrives at the formula (30).
\twocolumngrid


\end{document}